\documentclass[useAMS,usenatbib]{mn2e}
\usepackage{times,amssymb}

\newcommand{\kms}{km~s$^{-1}$}
\usepackage{graphics,epsfig}

\title[HI search in peculiar dwarfs ]{A search for HI in some peculiar faint dwarf galaxies}
\author[Begum \& Chengalur]
{
Ayesha Begum\thanks{E-mail:ayesha@ncra.tifr.res.in} and
Jayaram N. Chengalur
\\
\\
$^{1}$National Centre for Radio Astrophysics, Post Bag 3, Ganeshkhind, Pune 411 007, India\\
}
\begin{document}

\date{}


\maketitle

\label{firstpage}

\begin{abstract}

We present a deep Giant Metrewave Radio Telescope (GMRT) search for HI~21 
cm emission from three dwarf galaxies, viz.  POX~186, SC~24 and KKR 25. Based,
in part, on previous single dish HI observations, these galaxies have been 
classified as a BCD, a dwarf irregular and a transition galaxy 
respectively. However, in conflict with previous single dish detections,
we do not detect HI in SC~24 or KKR~25. We suggest that the previous 
single dish measurements were probably confused with the local galactic 
emission. In the case of POX~186, we confirm the previous non detection 
of HI but with substantially improved limits on its HI mass. Our derived 
upper limits on the HI mass of SC 24 and KKR 25 are similar to the typical
 HI mass limit for dwarf spheroidal galaxies, whereas in the case of 
POX~186, we find that its gas content is somewhat smaller than is typical of BCD galaxies.

\end{abstract}


\section{Introduction}

Here we present a search for HI in three peculiar dwarf galaxies,
POX~186, SC~24 and KKR~25. Based, in part, on previous, single 
dish HI observations, these galaxies have been classified as a blue
compact dwarf (BCD), a dwarf irregular  and a transition galaxy respectively. 

            POX~186 is an unusually compact BCD galaxy with a linear
size of only $\sim 300$~pc\footnote{ Following Guseva et al.~(2004) we 
assume a distance of 18.5~Mpc for this galaxy.}. HST observations (Corbin \& Vacca 2002) 
found an unusual asymmetry in the galaxy which they interpreted as a
tidal feature.  They hence argued for POX~186 being the result
of a recent ($<10^8$ yr) collision between two sub-galactic sized clumps,
and thus representative of a dwarf galaxy in formation. However this conclusion 
remains controversial as, from a study of ionized gas emission from 
the galaxy, Guseva et al.(2004) argued that the asymmetry in the morphology 
of POX 186 comes from a starburst-driven gaseous shell and not
from tidal arms. As signatures of tidal interactions are often most
prominent in HI, a deep HI image of POX~186  would be crucial in 
resolving this issue. While single dish observations did not detect  
any HI from the galaxy (Kunth et al. 1988), the limit on the HI mass of
the galaxy of $\sim 2.0 \times 10^7$ M$_\odot$ is somewhat higher 
than the expected HI mass  for the galaxy ($\sim 6 \times 10^6$ M$_\odot$, 
assuming M$\rm{_{HI}/L_B}$ $\sim 2$, which is a typical value for BCDs of luminosity
comparable to POX 186 $-$ e.g. Pustilnik et al. 2002).

KKR 25 (M${\rm{_B}}\sim-9.96$) has been classified as a  "transition" galaxy
 because of its significant HI content ($\sim 1.2\times 10^6 \rm{M_\odot}$ $-$ 
Huchtmeier et al. 2003) and the presence of a population of faint blue stars 
(Karachentsev et al. 2001). However, Grebel et al.(2003) point out that 
properties of KKR 25 are consistent with those of dwarf spheroidal (dSph) 
galaxies, as no H${\alpha}$ was detected from the galaxy and it also 
lies in the same region of the metallicity-luminosity plane as occupied 
by dwarf spheroidal galaxies. Based on the I magnitude of the tip of 
the red giant branch, Karachentsev et al. (2001) estimated the distance 
to KKR~25 to be 1.86 Mpc.

    The dwarf irregular galaxy SC~24 is the lowest luminous  member of the
Sculptor group and is also one of the faintest (M$\rm{_{B}}\sim-$8.39)
irregular galaxies known. This galaxy was discovered by C\^{o}te et al. (1997),
as a part of a survey of dwarf galaxies of the Sculptor group. Based on the 
distance-velocity relationship for the Sculptor group galaxies, Skillman et 
al.(2003) derived a distance of 1.66 Mpc to SC~24.

Our new Giant Metrewave Radio Telescope (GMRT) observations failed to 
detect HI in any of these galaxies. In what follows, we discuss our new 
GMRT observations, and the impact of our non detection of HI on the 
classification of these galaxies.

\section[]{Observations, analysis and results}
\label{sec:obs}

HI 21cm observations of KKR 25, SC 24 and POX 186 were conducted with the
GMRT (Swarup et al. 1991). The setup for the  observations is given
in Table~\ref{tab:obs}.

\begin{table}
\begin{center}
\caption{Parameters of the GMRT observations}
\label{tab:obs}
\hskip -0.5in
\begin{tabular}{lccc}
\hline
Parameters&KKR 25& SC 24&POX 186\\
\hline
\hline
RA (2000) & 16$^h$13$^m$47.6$^s$& 00$^h$36$^m$38.0$^s$& 13$^h$25$^m$48.6$^s$\\
DEC (2000) &  +${54}^{\circ} 22' 16''$&  $-{32}^{\circ} 34' 28''$&  $-{11}^{\circ} 36' 38''$\\
Observations date& 25 \& 26 Oct & 15 \& 27 June & 30 Nov$-$2 Dec \\
& 2001& 2003&  2002\\
Time on source (hrs) & 10 & 3 & 16 \\
Vel coverage (\kms)& -$45~-~$-256 & -343 $-$ 501& 1105 $-$ 1316\\
Vel resolution (\kms)& 1.6 & 6.6 & 1.6 \\
\hline
\end{tabular}
\end{center}
\end{table}

     The data were reduced in the usual way using standard tasks in classic AIPS.
The GMRT  has a hybrid configuration which simultaneously provides both high angular
resolution ($\sim 3^{''}$ if one uses baselines between the arm antennas) as well as
sensitivity to extended emission (from baselines between the antennas in the
central array). Data cubes were therefore made using various (u,v) cutoffs,
allowing a search for HI and continuum emission at various spatial resolutions
(given in Table~\ref{tab:result}).

        For all the galaxies, the data cubes were examined for line emission 
at a variety of
spectral resolutions $-$ in all cases no significant emission
was found. Besides visual inspection, the AIPS task SERCH was also used to search for
line emission in all  data cubes. No statistically significant feature was detected
in the cubes. To derive the final limit on the HI mass for those galaxies
with previous single dish HI detections, the data were smoothed to  the velocity 
width of the single dish spectrum. In particular, for KKR~25 a velocity 
width of 15 \kms was used (Huchtmeier et al. 2003), while velocity widths 
of 55 \kms  and 21 \kms were used for SC~24 (see Sect.~\ref{ssec:dis_sc24}). 
For POX~186, in the absence of a  previous single dish detection a 
velocity width of $\sim$20 \kms (a typical velocity width for such faint 
galaxies e.g. Begum \& Chengalur 2004, Begum et al. 2003) was used.
Finally, continuum images at $4^{''}\times3^{''}$ resolution  were also 
made for all the galaxies by averaging central 80 channels. No continuum 
was detected  in any galaxy. The derived 3$\sigma$ limits from the continuum
images are given in Table~\ref{tab:result}. 

        Table~\ref{tab:result}  summarizes our results for a representative
selection of spatial resolutions for three galaxies. Col.~(1) represents 
the galaxy name, (2)
the spatial resolution defined by the half-power beam width (HPBW) of the
synthesised beam, (3) RMS noise corresponding to this spatial resolution,
(4) the 5$\sigma$ upper limit on the HI mass (5) velocity resolution used
for deriving the limits ($\Delta$v) on the HI mass and (6) derived $3\sigma$ 
limit from the continuum image (RMS$_{\rm{cont}}$) .


\begin{table}
\begin{center}
\caption{Results of the GMRT observations}
\label{tab:result}
\begin{tabular}{|lcccccc|}
\hline
Galaxy&HPBW& RMS & M${\rm{_{HI}}}$& $\Delta$v & RMS$_{\rm{cont}}$  \\
&arcsec$^2$& (mJy)& $(10^{5}\rm{M_\odot})$&  (\kms)&(mJy)\\
\hline
\hline
KKR 25&49$\times44$&1.3&$<$0.8&15&0.75\\
\hline
SC 24&50$\times42$&1.2&$<$2.1&55&0.9\\
&50$\times40$&1.6&$<$1.1&21\\
\hline
POX 186&7$\times7$&0.3&$<$24.1&20&0.6\\
&4$\times3$&0.2&$<$16.1&20\\
\hline
\end{tabular}
\end{center}
\end{table}

\section[]{Discussion}
\label{sec:discuss}

The non detection of  HI with the GMRT in KKR~25 and SC~24, in conflict with the 
previous single dish detections of these galaxies, could be because 
the HI emission has resolved out by the GMRT. However, our past experience 
with the GMRT in successfully imaging dwarf galaxies with HI flux and optical sizes  
similar to SC 24 and KKR 25 (Begum et al. (2005) in preparation), makes such a possibility 
unlikely. Further, as discussed below, for both galaxies, a case can be made for 
confusion of emission from the galaxy with local HI. 


\subsection{KKR 25}
\label{ssec:dis_kkr25}

The Leiden/Dwingeloo survey of galactic neutral hydrogen (Hartman 1994) detects 
HI in the direction of KKR 25 in the same velocity range as detected by 
the single dish observations for this galaxy, it is hence likely that
the HI emission that has been associated with this galaxy actually
arises from local gas. The upper limit on the HI mass from the 
current GMRT observations is comparable to the limits obtained for 
typical dwarf spheroidal galaxies (Mateo 1998 and references therein).

\subsection{SC 24}
\label{ssec:dis_sc24}

There have been two separate claims of detection of HI in SC~24, viz. C\^{o}te et al. 
(1997) and Skillman et al. (2003).  The single dish flux measured  
by C\^{o}te et al. (1997) is  11.8 Jy \kms, with  a width at the 50\% level 
$\Delta  {\rm{V_{50}}}$$\sim$ 55.0 \kms. On the other hand, the HI flux 
estimated from the 
HIPASS database at the position of SC~24, is only 3.2 Jy \kms with 
$\Delta {\rm{V_{50}}}\sim 21~$\kms (Skillman et al. 2003).  

We note that in the case of the C\^{o}te et al. (1997) observations,
the signal to noise of the HI spectrum is poor and that there is a strong 
possibility of confusion with local HI. Indeed,
the HIPASS survey detects HI from the Magellanic stream in the direction 
of SC~24 at the radial velocity of the galaxy (Putman 2003). 

     If one goes by the current GMRT observations, the upper limit on
the HI mass of SC~24 is similar to the typical HI mass limits for dSph 
galaxies (see Mateo 1998 and references therein). However the optical 
appearance of the galaxy suggest that it is a dwarf irregular galaxy 
(Karachentsev et al. 2004).  Further, unlike other member galaxies of the Sculptor 
group, the  HST observations failed to resolve this galaxy  into stars 
(Karachentsev, I. D., private communication).  Hence, SC 24 is more 
likely to be a distant galaxy. 

\subsection{POX 186}
\label{ssec:dis_pox186}

      Our GMRT observations for POX 186 confirms the previous single dish non detection,
albeit with a much better limit on the HI mass. The upper limit on the HI mass of the galaxy 
computed\footnote{The velocity width and confidence limit used to calculate 
the upper limit on the HI mass for POX 186 is not specified in the paper}
from the single dish observations is $\sim2\times10^7\rm{M}_\odot$ (Kunth et al. 1988),
whereas the 5$\sigma$ upper limit on the HI mass derived from our interferometric
observations is $2.4\times10^6 {\rm{M}}_\odot$. Since HI in dwarf galaxies typically extends to 
$\sim$ 2 times the Holmberg diameter (e.g. Hunter 1997), and the angular size of
the optical emission is $\sim3^{\prime\prime}$, (Corbin \& Vacca 2002), the upper 
limit we quote is derived from the $7^{\prime\prime}\times7^{\prime\prime}$ 
resolution data cube, and for a velocity width of $20$~\kms. HST observations 
of POX~186 found a young OB stellar cluster with an estimated age of $\sim10^6-10^7$~yr 
(Corbin \& Vacca, 2002). We searched for HI emission near the location of this cluster 
at the highest angular resolution of our data (4$''\times3''$) but did not find any 
statistically significant emission. Our 5$\sigma$ upper limit (again for a velocity 
width of $20$ km/s) is $1.6\times10^6 {\rm{M}}_\odot$. The absence of HI in 
POX 186 is some-what puzzling, as these young stars must be associated with 
some neutral HI. From a statistical study of a sample of BCDs in various environments, 
Pustilnik et al. (2002) found that BCDs in  voids have a higher M$_{\rm HI}$/L$_{\rm B}$
than those in higher density regions and that there is also a trend for increasing 
M$_{\rm HI}$/L$_{\rm B}$ with  decreasing L$\rm{_B}$.  From these correlations,
the expected M$_{\rm HI}$/L$_{\rm B}$ for POX~186 is $\sim 2$. Hence, our derived 
upper limit on the HI mass of POX 186, (which corresponds to M$_{\rm HI}$/L$_{\rm B}
\lesssim 0.8$) means that the galaxy has a somewhat  smaller M$_{\rm HI}$/L$_{\rm B}$ 
than is typical for BCDs.  However it should be noted that the scatter in
M$_{\rm HI}$/L$_{\rm B}$ for BCDs at a given luminosity is large, and further
that the correlations found by Pustilnik et al. (2002) were computed for 
galaxies brighter than M$\rm{_B}\sim -16.0$ mag, and it is not clear whether 
the fainter BCDs follow a same trend as the brighter ones. Given the small size 
of the galaxy and the fact that a significant amount  of ionized gas has been 
detected in POX~186 (Guseva et al. 2004), it is possible that 
a sizeable fraction of HI in the galaxy has been ionized by the recent burst of
star formation.

\section[]{Conclusions}

   To conclude, despite a deep search, we find no HI emission associated with 
POX~186, SC~24 and KKR~25. The non-detection of HI in SC 24 and KKR 25 
suggests that previous  single dish measurements were affected by confusion with 
the galactic emission. Our stringent limits on the HI mass of KKR 25 indicate 
that it is a normal dwarf  spheroidal galaxy, whereas SC 24 is more likely to 
be a distant galaxy.  In the case of POX~186, the derived HI mass limit is 
somewhat smaller than is typical for BCDs.

\section*{Acknowledgments}

We are grateful to I. D. Karachentsev for
providing important information about SC 24 which helped us in 
interpreting our results. The GMRT is operated
by the National Center for Radio Astrophysics of the Tata Institute
of Fundamental Research.

\end{document}